\documentclass[twocolumn,aps,prb,reprint,groupedaddress]{revtex4}
\usepackage{graphicx}
\usepackage{color}
\usepackage{amsmath}

\usepackage{hyperref}
\hypersetup{
  colorlinks   = true, %Colours links instead of ugly boxes
  urlcolor     = blue, %Colour for external hyperlinks
  linkcolor    = blue, %Colour of internal links
  citecolor   = blue %Colour of citations
}

\newcommand{\be}{\begin{equation}}
\newcommand{\ee}{\end{equation}}

\newcommand{\bea}{\begin{eqnarray}}
\newcommand{\eea}{\end{eqnarray}}

\newcommand{\p}{\partial}

\newcommand{\la}{\langle}
\newcommand{\ra}{\rangle}

\newcommand{\lp}{\left(}
\newcommand{\rp}{\right)}

\renewcommand{\vec}[1]{{\bf #1}}
\renewcommand{\hat}[1]{{\widehat #1}}

\newcommand{\tr}{{\rm Tr}}

\begin{document}
\title{Electron-Electron Interactions and Plasmon Dispersion in Graphene}
%\author{people}
%\affiliation{places}
\author{L. S. Levitov}
\affiliation{Massachusetts Institute of Technology, 77 Massachusetts Avenue, Cambridge MA 02139, USA}
\author{A. V. Shtyk and M. V. Feigelman}
\affiliation{L. D. Landau Institute for Theoretical Physics, Kosygin Street 2, Moscow, 119334 Russia}
\affiliation{Moscow Institute of Physics and Technology, Institutsky Pereulok 9, Dolgoprudny, 141700 Russia}

%\email{levitov@mit.edu}

\begin{abstract}
Plasmons in two-dimensional electron
systems with nonparabolic bands, such as graphene, feature strong dependence 
on electron-electron interactions. We use a many-body approach to relate plasmon dispersion at long wavelengths
to Landau Fermi-liquid interactions and quasiparticle velocity. 
An identical renormalization is 
shown to arise 
for the magnetoplasmon resonance. For a model with $N\gg 1$ fermion species, this approach predicts a
power-law dependence for plasmon frequency vs carrier concentration, valid in a wide range of doping densities, both high and low. 
%Gate tunability of plasmons in graphene can be used to directly probe the effects of electron-electron interaction.
Gate tunability of plasmons in graphene can be
exploited to directly probe the effects of electron-electron interaction.
%
%Plasmons in two-dimensional systems with non-parabolic band, such as graphene,  feature strong dependence on electron-electron interactions. We use a many-body approach to relate plasmon dispersion at long wavelengths to the Landau's Fermi-liquid interaction parameters and quasiparticle velocity. This general relation is illustrated using a model with $N$ fermion species, giving a power-law dependence for plasmon frequency vs. carrier density at leading order in large $N$. An identical renormalization is predicted for magnetoplasmon resonance. We propose that gate-tunability of plasmons in graphene can be used to directly probe the effects of electron-electron interaction.
\end{abstract}


\maketitle
Plasmonics has emerged recently as an active direction in graphene  research.\cite{wunsch06,hwang1,polini2009,dassarma09,mishchenko2010} %,abedinpour2011} 
Surface plasmons in 2D electron systems are propagating charge density waves in which collective dynamics of clouds of charge is mediated by electric field in 3D.\cite{giuliani2005} 
%Collective density oscillations in electron liquid--plasmons--are of great interest because they combine in a unique way the properties of photons and charge carriers. Plasmon propagation velocity and wavelength values lie between the values typical for charge carriers and light of the same frequency\cite{giuliani2005}.  
The dual matter-field nature of plasmons is a key ingredient for many interesting and important phenomena.\cite{theis80,chaplik85} 
%However, even the best plasmonic materials available to date (prior to the recent advent of graphene), such as noble metals and related materials, suffer from a number of drawbacks. In particular, they are only slightly tunable and exhibit large Ohmic losses. 
Plasmons in graphene display a range of potentially useful properties, such as low Ohmic losses, a high degree of field confinement, and gate-tunability.
%, which make these excitations of great interest for optoelectronics.
\cite{bonaccorso,koppens2011} Gate-tunability of plasmons in graphene was demonstrated recently.\cite{ju2011,chen2012}

The goal of this article is to 
%relate the density dependence of plasmons with the interactions 
investigate the density dependence of plasmons and relate it to the interaction effects in the electron system. 
%Plasmon dispersion depends on carrier density due to several effects. It takes simplest form in the limit of weak electron-electron interactions
The dependence of plasmon dispersion on carrier density arises due to several effects. It takes on the simplest form in the limit of weak electron-electron interactions,\cite{wunsch06,hwang1,polini2009,dassarma09}
\be\label{eq:noninteracting}
%\omega^2=\frac{e^2 E_F}{\pi\hbar^2\kappa}q
\omega^2=\frac{2e^2 E_F}{\kappa\hbar^2}q
\ee
where $E_F\sim n^{1/2}$ is the Fermi energy of noninteracting massless Dirac particles, and $n$ is carrier density. Here $\kappa$ is an effective dielectric constant of the substrate and a long-wavelength limit is assumed, $q\ll p_F $. %Interestingly, 
%the density dependence of the 
%plasmon dispersion is also affected by interactions. Renormalization of the dispersion relation $\omega(q)$, Eq.(\ref{eq:noninteracting}), 
%due to the effects of electron-electron interaction was 
Furthermore, plasmon dispersion features strong dependence on interactions. Renormalization of the dispersion relation, Eq.(\ref{eq:noninteracting}), due to electron-electron interactions was
predicted in Ref.\onlinecite{polini2009},  where perturbation expansion in a weak fine structure parameter $\alpha=e^2/\hbar v$ was employed. The results of Ref.\onlinecite{polini2009} point to an interesting possibility to directly probe the effects of 
%electron 
interactions by measuring plasmon dispersion relation. However, 
%because of 
strong interactions in graphene, $\alpha\sim 2.5$, 
%the weak coupling approximation is unrealistic.
render the weak coupling approximation unreliable.
% cannot serve as a good model, and a different approach is required.

Acknowledging the difficulty of modelling the strong-coupling regime, it is beneficial to 
%approach the problem at a somewhat more general angle.
adopt a somewhat more general approach.
Rather than attempting to make predictions based on a specific microscopic models, one can ask if a relation between 
the plasmon dispersion and some other fundamental characteristics of the system can be established.
Below we point out that such a relation arises naturally from the Landau theory of  Fermi liquids.\cite{lifshitz_pitaevskii} This theory affords a general, model-independent framework to describe systems of strongly interacting fermions at degeneracy.  The effects of interactions are 
%described by 
encoded in the Landau parameters, representing a ``genetic code'' of the Fermi liquid (FL). 
%Their values
The parameter values can in principle be predicted from perturbation theory if interactions are weak. For systems with strong interactions, however,
%%, however,
the most reliable way to obtain the Landau parameters is to use their relation with experimentally measurable quantities, such as compressibility, heat capacity, spin susceptibility and dispersion of collective excitations.

Our many-body analysis upholds the conventional square-root dependence $\omega\sim q^{1/2}$. We show that all the effects of interactions are accumulated in the prefactor,
\be\label{eq:plasmon_dispersion0}
\omega^2={Y} \lambda q
,\quad
{Y} =(1+F_1)v
,\quad
\lambda=\frac{Ne^2p_F }{2\kappa\hbar^2}
\ee
where $p_F $ is Fermi momentum, $N=4$ is the number of spin/valley flavors, and a long-wavelength limit is assumed, $q\ll p_F $. Here $F_1$ is the Landau interaction harmonic with $m=1$, and $v$ is the Fermi velocity renormalized by interactions.
The quantity $\lambda$ in Eq.(\ref{eq:plasmon_dispersion0}) has units of frequency and depends only on the fundamental constants and carrier density via $p_F $. 
%The dielectric constant $\kappa$ may have a $q$ dependence describing image charges arising due to conducting boundaries such as gates. 
In some cases the dielectric constant $\kappa$ may feature an essential $q$ dependence. 
%This is the case, 
In particular, when image charges arise due to conducting boundaries and gates, a simple model yields\cite{chaplik72} $\kappa(q)=\frac12(\kappa_1+\kappa_2\coth(qd))$, giving an acoustic plasmon dispersion $\omega\sim q$.
%The quantity $\lambda$ in Eq.(\ref{eq:plasmon_dispersion0}) has units of frequency and depends only on the fundamental constants and carrier density via $p_F $. 

Magnetic field 
%turns
alters the behavior, turning the gapless plasmon mode into a gapped mode. The magnetoplasmon dispersion relation obtained by adding Lorentz force to the FL dynamics takes on the following form:
\be\label{eq:magnetoplasmon_dispersion}
\omega^2_B(q)=\omega_0^2(q)+{Y}^2\lp eB/cp_F\rp^2,\quad q r_c\ll 1,
\ee
where $\omega_0(q)$ is plasmon dispersion at $B=0$ given by Eq.(\ref{eq:plasmon_dispersion0}), and $r_c$ is the cyclotron radius. The dispersion relation becomes more complex at $q r_c\sim 1$ due to the presence of Bernstein modes.\cite{bernstein}
The size of the gap at $q=0$ scales linearly with $B$, 
%with the prefactor that varies inversely with $p_F\sim n^{1/2}$.
with a density dependent prefactor.
Notably, the magnetoplasmon dependence on the interactions is described by the same combination ${Y}=(1+F_1)v$ as that appearing in Eq.(\ref{eq:plasmon_dispersion0}).

The quantity ${Y}$ describes the interaction dependence of plasmon dispersion. Measuring  it as a function of carrier density 
%can shed light on the role of electron-electron interactions
can be used to determine the electron-electron interaction strength
in the system. This behavior is in sharp departure with that for plasmons in two-dimensional systems with parabolic band dispersion, where Galilean invariance leads to an identity for Landau parameters, $(1+F_1)v=v_0$,\cite{lifshitz_pitaevskii} where $v_0=m p_F $ is Fermi velocity of noninteracting particles at the same density 
%[this is discussed in more detail
[see discussion in Sec.\ref{sec:galilean}]. As a result, the value $(1+F_1)v$ is independent of interactions, leading to the ``universal'' long-wavelength plasmon dispersion in the parabolic case, $\omega^2_0(q)=\frac{2\pi e^2n}{m\kappa(q)}q$.
%at $B=0$ and 
Similarly, at a finite magnetic field, Galilean invariance leads to a simple result for long-wavelength magnetoplasmons,
$\omega^2_B(q)=\omega_0^2(q)+\omega_c^2$, 
%for magnetoplasmon (
where $m$ is unrenormalized band mass, $\omega_c=eB/mc$ is the cyclotron frequency
%)
and $q r_c\ll 1$. These dependences carry no information on the quantum effects and neither on the interactions.

The situation is quite different in systems with nonparabolic dispersion, such as graphene.
The density dependence in ${Y}$ arises because the values $v$ and $F_1$ 
%depend on the phase volume of the states in the conduction band contributing to renormalization. The energies of these states span the domain from graphene bandwidth down to the Fermi energy, which is gate-tunable.
are renormalized in an essentially different way.
%To illustrate this point, 
As an illustration, we analyze the limit of a large number of spin/valley flavors, $N\gg1$, using a renormalization group (RG) approach. In this case, as we will see, Eq.(\ref{eq:plasmon_dispersion0}) yields a power-law dependence on carrier density, %giving
\be
\omega^2\sim A n^{(1-\beta)/2} q
.
\ee
Here the exponent $\beta$ is identical to  that found from one-loop RG for velocity renormalization, $\beta=\frac{8}{N\pi^2}$,
\cite{gonzalez1994,vafek2007,son2007} and the prefactor is $A\sim v_0 \frac{e^2}{\hbar\kappa} a^{-\beta}$, with $a\approx 0.142\,{\rm nm}$ the carbon spacing.
For a non-interacting system, $\beta=0$. Crucially, the power law $n^{(1-\beta)/2}$ describes the dependence on carrier density not only near charge neutrality but also for all accessible $n$ values. 

Measurements of the density dependence of a plasmon 
%resonance were 
resonance in graphene ribbons were reported in Ref.\onlinecite{ju2011}. The observed dependence approximately follows the relation $\omega^2\propto q$, with the prefactor exhibiting an approximately linear dependence on $n^{1/2}$. 
%However, the range of densities in which the dispersion was measured was not wide enough to provide a clear distinction 
However, the limited range of densities in which the dispersion was measured, as well as possible corrections due to the finite width of the ribbons, made it challenging to distinguish between $\beta=0$ and $\beta\ne 0$.
An attempt to  experimentally determine the RG scaling 
exponents directly from transport measurements was made recently in Ref.\onlinecite{elias2011}. In this work, a systematic variation of the period of quantum oscillations with carrier density was interpreted in terms of Fermi velocity renormalization,  giving a value $\beta=0.5$-$0.55$. This value is considerably larger than the one-loop RG result, $\beta=\frac8{\pi^2 N}\approx 0.2$.
\cite{gonzalez1994,vafek2007,son2007} This discrepancy is not yet understood.
% For the plots we will use a conservative value $\beta=0.3$. 

We note parenthetically that the interaction effects are not expected to vanish in graphene bilayer despite the parabolic character of its band dispersion. Electronic states in graphene bilayer are Dirac-like rather than Schr\"odinger like, and hence do not admit Galilean transformation. For plasmons in this material we therefore expect a behavior similar to that in materials with nonparabolic band, described by Eqs.(\ref{eq:plasmon_dispersion0}),(\ref{eq:magnetoplasmon_dispersion}).

\section{Microscopic Fermi-liquid analysis}
\label{sec1}
The goal of this section is to relate plasmon dispersion with the standard quantities such as Landau FL interactions and renormalized velocity. The analysis proceeds by standard steps via resumming the ladder contributions to the dynamical polarization function which account for the quasiparticle dynamics in Landau's FL framework. In doing so, we keep $\omega$ and $q$ small but finite, as appropriate for a plasmon dispersion analysis. This leads to a polarization response, $\Pi(q,\omega)\sim q^2/\omega^2$, describing plasmon excitations in the low-frequency and long-wavelength domain, $\omega\ll E_F$, $q\ll p_F$.

Charge carriers in graphene single layer are described by the Hamiltonian for $N=4$ species of massless Dirac particles. In second-quantized representation the Hamiltonian reads
\bea
&&
\mathcal{H}=  \sum_{\vec p,i}  \psi^\dag_{\vec p,i} v_0\boldsymbol{\sigma}\vec p  \psi_{\vec p,i}  + \mathcal{H}_{\rm el-el}, 
%\mathop{\sum_{\vec q, \vec k} }_{\vec{k'},i,j}V_{ij}(\vec q) \psi^\dag_{\vec k + \vec q, i} \psi^\dag_{\vec{k'} - \vec q,j} \psi_{\vec {k'},j} \psi_{\vec k, i} 
\label{eq:hamiltonian}
\\\label{eq:interaction}
&&
\mathcal{H}_{\rm el-el} = \frac12\sum_{\vec q, \vec p,\vec{p'},i,j}V(\vec q) \psi^\dag_{\vec p + \vec q, i} \psi^\dag_{\vec p' - \vec q,j} \psi_{\vec p',j} \psi_{\vec p, i} 
,
\eea
where $i,j=1...N$, $v_0\approx10^6{\rm m/s}$ is unrenormalized Fermi velocity, and $V(\vec q)=2\pi e^2/|\vec q|\kappa$ is the Coulomb interaction with the dielectric constant $\kappa$ describing screening by the substrate.
Here $\psi_{\vec p,i}$ is a two-component spinor describing the wave-function amplitude on the two sublattices of the graphene crystal lattice. The amplitudes associated with the two sublattices are usually referred to as pseudospin up and down components, with the (pseudo)spin-$1/2$ Pauli matrices in Eq.(\ref{eq:hamiltonian}) acting on (pseudo)spinors $\psi_{\vec p,i}$.

\begin{figure}
\includegraphics[width=1\linewidth]{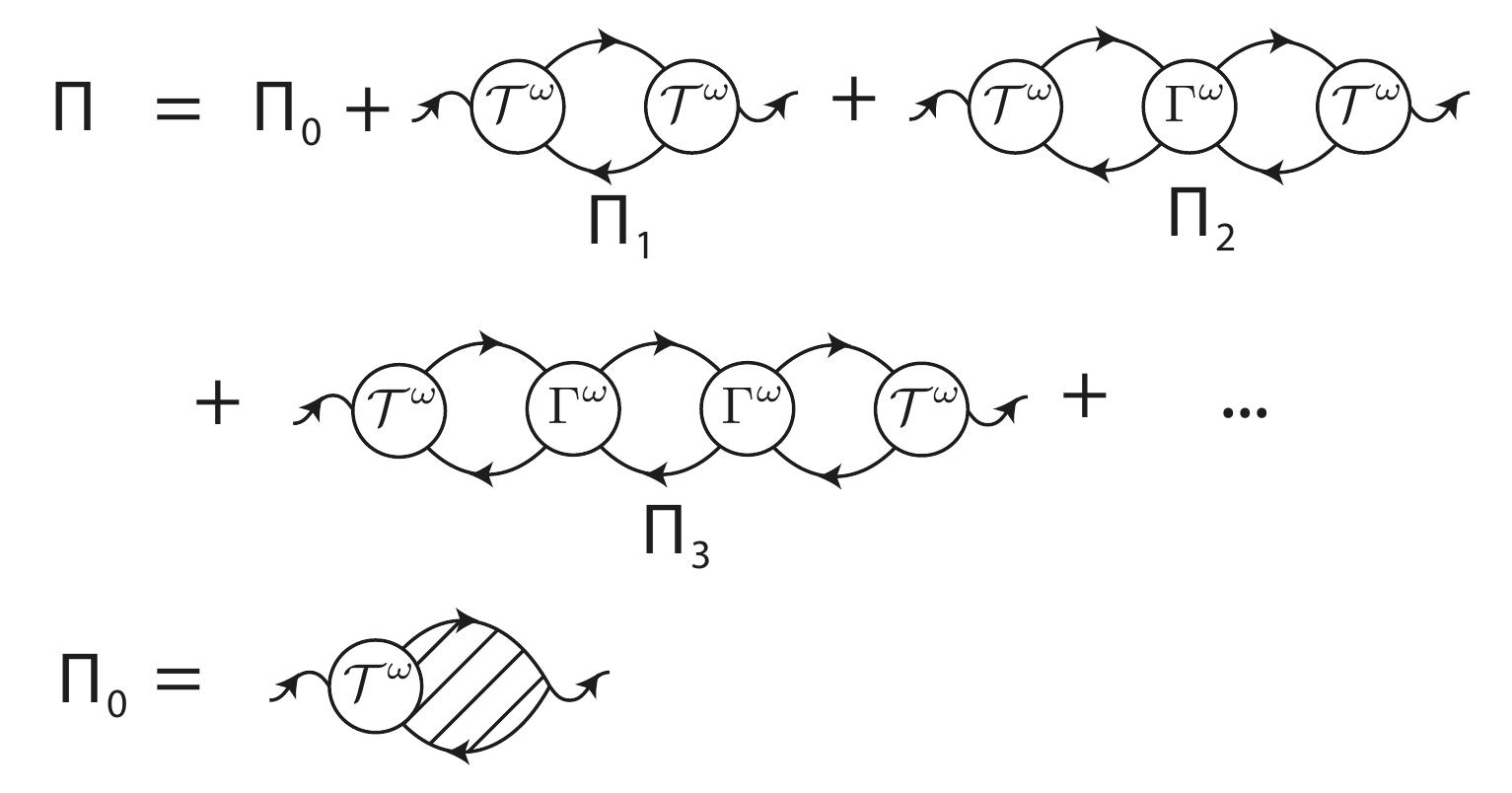}
\caption{Resummed Feynman graphs for the polarization operator $\Pi(\vec q,\omega)$. The non-quasiparticle contribution $\Pi_0(\vec q,\omega)$ and the FL ladder $\sum_{n\geq1}\Pi_n(\vec q,\omega)$ are shown. Only the contributions $\Pi_1(\vec q,\omega)$ and $\Pi_2(\vec q,\omega)$ contribute to the low-energy plasmon dispersion, see text.}
\label{fig:pi}
\end{figure}

Plasmons are collective excitations of 2D electrons coupled by the electric field in 3D. They can be  described microscopically using the density correlation function
%of Fermi liquid it arise as the pole of density-density correlation function
%
\be
\label{eq:K}
%K(\vec q,\omega)=\sum_{\vec p}\left<\psi^+_\alpha(\varepsilon_-,\vec p_-)\psi_\alpha\psi^+_\beta\psi_\beta\right>
K(\vec q,\omega)=i\int dt \la [\rho_{\vec q}(t),\rho_{\vec q}(t_0)] \ra e^{i\omega (t-t_0)}
\ee
where  $\rho_{\vec q}(t)=\sum_{\vec p,i}\psi^\dagger_{\vec p,i}(t)\psi_{\vec p +\vec q,i}(t)$ are  Fourier harmonics of the total electron density.
The quantity $K$ is expressed in a  standard fashion\cite{lifshitz_pitaevskii} through geometric series involving the polarization function $\Pi(\vec q,\omega)$ defined as the irreducible density-density correlator, 
\be\label{eq:vareps_zero}
K(\vec q,\omega)=\frac{\Pi(\vec q,\omega)}{\tilde{\kappa}(\vec q,\omega)}
,\quad
\tilde{\kappa}(\vec q,\omega)=1-V(\vec q)\Pi(\vec q,\omega)
\ee
Zeros of the dynamical screening function  $\tilde\kappa(\vec q,\omega)$ give the poles of $K$, defining plasmon dispersion.
To obtain the dispersion from the condition $\tilde\kappa(\vec q,\omega)=0$ we need 
%to gain information
an input
on $\Pi(\vec q,\omega)$ from a microscopic approach. In the long-wavelength limit, $q\ll p_F $, $\omega\ll E_F$, the behavior of the quantity $\Pi(\vec q,\omega)$ is dominated by excitations near the Fermi surface, which can be described in the FL framework.

%The microscopic approach which forms the basis of the FL phenomenology involves several standard steps.  At a first step, we isolate the contribution due to the quasiparticle pole of the 
The microscopic approach used to justify the FL picture involves several standard steps. We start, as usual, by isolating a quasiparticle pole contribution to the
electron Greens function $G(x-x')=-i\left<\psi(x)\psi^\dagger(x')\right>$ near the Fermi surface,
\be\label{eq:G_pole}
G(\epsilon,\vec p)=
G^{\rm (reg)}(\epsilon,\vec p) + G^{\rm (sing)}(\epsilon,\vec p)
.
\ee
%, and $\alpha$, $\beta$ are flovor (spin/valley) indices.  
The first term is a regular part of the Greens function
% arising due to electron-electron interactions
behaving as a smooth function near the Fermi level. The second term is a singular contribution describing quasiparticles,
\be
G^{\rm (sing)}(\epsilon,\vec p)=\frac{Z}{i\epsilon-\xi(p)+i\gamma \,\mathrm{sgn} \epsilon}
.
\ee
Here $Z$ is a quasiparticle residue, $\gamma$ is a quasiparticle decay rate, and $\xi(p)=v(p-p_F)$ is a quasiparticle energy dispersion, with $v$ the renormalized velocity.

This general 
%relation 
discussion
can be specialized to the case of graphene as follows. The Green's function for electrons in graphene has a $2\times 2$ matrix pseudospin structure. By projecting on the conduction and valence bands, it can be represented as
\be
\hat G(\varepsilon,\vec p)
%=-i\left<\psi_\alpha\psi^\dagger_\beta\right>
=G_<(\varepsilon, \vec p)\hat{P}_<+G_>(\varepsilon, \vec p)\hat{P}_>
\ee
where $\hat{P}_{>(<)}=(1\pm\boldsymbol{\sigma} \vec e_p)/2$ are projectors for the two bands (here $\vec e_p$ is a unit vector in the direction of momentum $\vec p$).
The quasiparticle excitations with low energies,  which govern the  low-frequency and long-wavelength response, reside near the Fermi level.
Without loss of generality, we assume n-type doping, so that the Fermi level lies in the upper band, $E_F>0$. In this case,  excitations from the lower band do not appear explicitly in the FL theory and lead only to renormalization of various parameters such as the  effective interactions and the quasiparticle velocity. The quasiparticle pole in Eq.(\ref{eq:G_pole}) therefore arises only from the upper-band contribution $G_>$ whereas the lower-band contribution $G_<$ can be absorbed into the regular part $G^{\rm (reg)}$. Below the subscripts $>$ and $<$ will be omitted for brevity.

\begin{figure}
\includegraphics[width=1\linewidth]{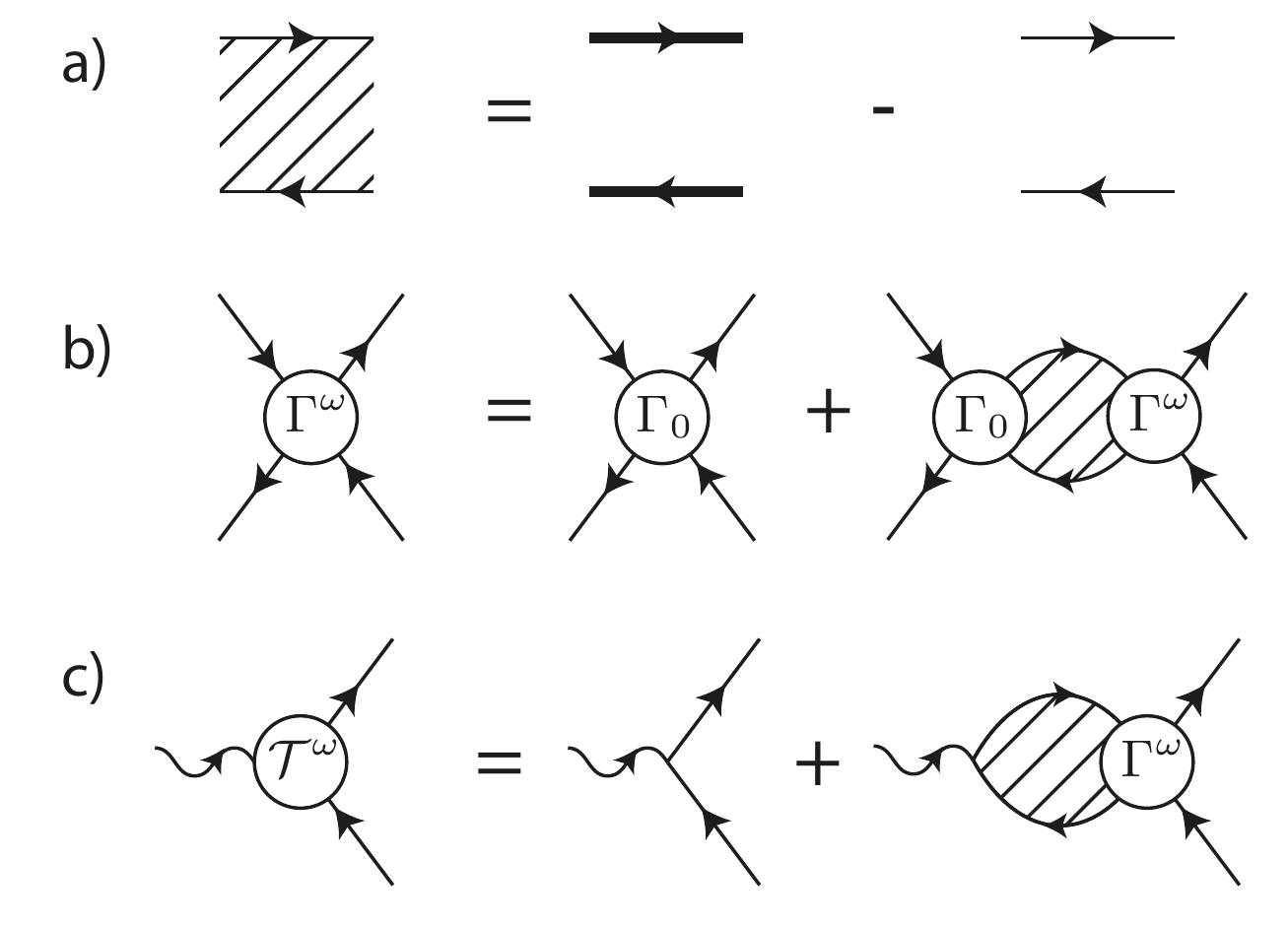}
\caption{
Feynman graphs for the ``quasiparticle-irreducible'' quantities $\mathcal{T}^\omega,\,\Gamma^\omega$. 
%Bold lines represent a full Green's function while the thin lines corresponds to
The bold lines represent a full Green's function, the thin lines represent
a singular part $G^{\rm (sing)}$. 
%The interaction vertex  $\Gamma_0$ represents the bare interaction
The vertex  $\Gamma_0$ represents the conventional irreducible vertex, whereas  hatched blocks represent products of two Green's 
%functions save for the 
functions, with the
contributions $G^{\rm (sing)}G^{\rm (sing)}$ which give a non-analytical behavior at small $\omega,\vec q$, 
%see text. 
taken out (see text).
}
\label{fig:gamma_t}
\end{figure}

The next step, which is key for understanding the role of low-energy excitations, is the analysis of the polarization function $\Pi(q,\omega)$ at small $\omega$ and $q$. This is done by identifying the contributions due to pairs of Greens functions with proximal poles (the ``dangerous'' two-particle crosssections),\cite{lifshitz_pitaevskii} which we write symbolically as $G^{\rm (sing)}G^{\rm (sing)}\sim Z^2\frac{\vec v\vec k}{\omega-\vec v\vec k}$. One can represent $\Pi(q,\omega)$ as a sum of terms with different numbers of such  contributions,
\bea\label{eq:P0+P1+P2}
&&\Pi(q,\omega)=\Pi_0(q,\omega)+\Pi_1(q,\omega)+\Pi_2(q,\omega)+...
\\\nonumber
&&
\Pi_1(q,\omega)=\mathcal{T}^\omega GG\mathcal{T}^\omega
,\quad
\Pi_2(q,\omega)=\mathcal{T}^\omega GG\Gamma^\omega GG\mathcal{T}^\omega
 ...
\eea
The corresponding graphs are shown in Fig.\ref{fig:pi}. Here we introduced so-called quasiparticle-irreducible quantities: the renormalized scalar vertex $\mathcal{T}^\omega$  and the two-particle scattering vertex $\Gamma^\omega$ (see Fig.\ref{fig:gamma_t} b,c).  These quantities absorb all non-quasiparticle contributions in the upper band as well as the inter-band 
processes and 
the contribution of the states in the lower band. 

We recall that the quasiparticle-irreducible quantities are distinct from the conventional irreducible quantities defined as sums of Feynman graphs that  cannot be split in two by removing two electron lines.\cite{lifshitz_pitaevskii} For example, the quasiparticle-irreducible vertex $\Gamma^\omega$ is obtained by summing all kinds of graphs except the ones with dangerous cross-sections. The vertex $\Gamma^\omega$ ($\mathcal{T}^\omega$) can be obtained from the conventional irreducible vertex $\Gamma_0$ ($\mathcal{T}_0$) by the resummation procedure pictured in Fig.\ref{fig:gamma_t}, where the hatched blocks represent contributions due to pairs of Greens functions save for $G^{\rm (sing)}G^{\rm (sing)}$.

To analyze the dependence on $\omega$ and $\vec q$ in the long-wavelength limit, 
%we exercise caution and employ
caution must be exercised by employing 
the quantities $\mathcal{T}^\omega$ and  $\Gamma^\omega$ taken at small but nonzero frequency and momentum values. We therefore adopt an approach similar to that used in Ref.\onlinecite{luttinger}: our quasiparticle-irreducible quantities correspond to Luttinger's $\omega$-quantities which are taken at finite $\omega$ and $\vec q$. They are distinct from the conventional $\omega$-quantities\cite{lifshitz_pitaevskii} obtained in the limit $\omega,q\to 0,\,(\omega\gg v_Fq)$. This distinction, however, turns out to be  inessential: Luttinger's $\omega$-quantities reproduce the conventional $\omega$-quantites in the limit $\omega,q\to 0$, which can be taken in arbitrary order since dangerous cross-sections were left out of the definition of $\mathcal{T}^\omega$ and $\Gamma^\omega$.

Proceeding with the analysis, we note that the dependence on $\omega$ and $q$ is very different for $\Pi_0(q,\omega)$ and $\Pi_{n\ge 1}(q,\omega)$. We will first analyze the  contribution $\Pi_0(q,\omega)$. This quantity does not contain dangerous cross-sections which can generate a nonanalytic behavior at small $\omega$ and $\vec q$. Taking $\Pi_0(q,\omega)$ to be analytic, we can represent it as
\be
\Pi_0(\vec q,\omega)=A(\omega)+B(\omega)\left(\frac{q}{p_F}\right)^2+...
\ee
where $A(\omega)$ and $B(\omega)$ are regular functions. 
%To realize that $f_1(\omega)=0$ we 
Further, we recall that gauge invariance prohibits any physical response to spatially uniform time-dependent scalar field. Applying this to the full polarization function, Eq.(\ref{eq:P0+P1+P2}), we 
%conclude
see that setting $q=0$ yields $\Pi(\omega,0)=0$. Also, since the contributions of the dangerous cross-sections $GG$ vanish at $q=0$, all the quantities $\Pi_{n\ge 1}(q,\omega)$ do so. We therefore conclude that the function $A(\omega)$ vanishes, leaving us with $\Pi_0(\vec q,\omega)=(q/p_F)^2B(\omega)+O(q^4)$. This gives an effective $q$-dependent permittivity
\be
\label{eq:kappa}
\tilde\kappa(q,\omega)=1-V(q)\Pi_0(q,\omega).
\ee
%
%\addLL{(power ${}^{-1}$ removed)}
The second term in Eq.(\ref{eq:kappa}) may henceforth be ignored in the long-wavelength limit. Indeed, since $V(q)=2\pi e^2/\kappa q$ in 2D, 
%the effective long-wavelength permittivity is equal to unity and may henceforth be ignored in the limit $q/p_F\to 0$. 
whereas $\Pi_0\propto q^2$, the quantity $\tilde{\kappa}$ equals unity in the limit $q/p_F \to 0$. 

%We note parenthetically that a very different 
It is instructive to compare 
%the
this behavior of $\tilde\kappa(q,\omega)$ with that arising for $q\gg p_F$. In this case the effects of finite doping are negligible and we can estimate the polarization function using the result obtained for massless Dirac particles at zero doping, 
\be\label{eq:Pi1}
\Pi(q,\omega)=-\frac{N}{16}\frac{q^2}{\sqrt{q^2v^2-\omega^2}}
,
\ee
where $N=4$ is the number of spin/valley flavors. In the limit $qv\gg \omega$, we obtain a well known 
%$q$-independent result for
renormalized permittivity
\be
\tilde\kappa(q,\omega)=1+\frac{\pi N\alpha}8
\ee
This $q$-independent expression describes the effect of intraband polarization in undoped graphene. We stress, however, 
%that while the above expression is different from unity, it is 
that while $\tilde{\kappa}>1$, the above expression is
obtained for $q$ and $\omega$ 
%which 
values which are not relevant for plasmon excitations. 
%In fact, 
This is so because plasmons do not exist for such $\Pi(\vec q,\omega)$, as the plasmon dispersion terminates for $q\gtrsim p_F $. 
%At the same time, the permittivity found in the long-wavelength limit
In contrast, the permittivity in Eq.(\ref{eq:kappa}), evaluated in the long-wavelength limit relevant for plasmons,
$q\ll p_F$, $\omega\ll E_F$, 
%which is relevant for plasmons, 
equals unity.

Next, we proceed with the analysis of the remaining terms, $\Pi_{n\ge 1}(q,\omega)$, which 
%contribute 
give a leading contribution to the low-energy plasmon dispersion.
This is the part of polarization which 
%account for
depends on the quasiparticle contributions. The corresponding Feynman graphs are given by ladder with rungs consisting of two quasiparticle lines separated by vertex parts, as shown on a Fig.\ref{fig:pi}. This gives geometric series that can be easily summed up:
% , where, owing to the fact that $\omega\gg v_Fq$, we keep only two rungs depicted on Fig. X
%
\be\label{eq:pi_interacting}
\Pi(\vec q,\omega)=
%\oint \frac{d\theta}{2\pi}\oint \frac{d\theta'}{2\pi}
N\oint\frac{d\theta}{2\pi}\mathcal{T}^\omega\lp \frac{\nu Z^2}{\omega-\vec q\vec v(1+\hat F)}\vec q\vec v\mathcal{T}^\omega\rp
%,\quad
% \hat F f(\theta)=\nu Z^2\oint\frac{d\theta'}{2\pi}\Gamma^\omega(\omega,\vec q,\theta,\theta')f(\theta')
,
\ee
where $\hat{F}$ is an integral operator
\be\label{eq:F=Gamma}
\hat F f(\theta)=\nu Z^2\oint\frac{d\theta'}{2\pi}\Gamma^\omega(\omega,\vec q,\theta,\theta')f(\theta')
.
\ee
Here $\nu=p_F /(2\pi\hbar^2 v)$ is 
%density of states per flavor,  $\theta$ is an angle between $\vec p$
the density of states per flavor and $\theta$($\theta'$) is an angle between $\vec p$($\vec p'$) and $\vec q$. 
%$\hat{F}$ is an integral operator and $\theta$ is an angle between $\vec p$ and $\vec q$.
%parameterizing the direction of $\vec p$ while without loss of generality $\vec q$ may be chosen to point in $\theta=0$ direction. 
For zero external momentum $\vec q\rightarrow0$ the kernel of the operator $\hat F$ depends only on the angle between $\vec p$ and $\vec p'$. 
%In fact, we will need only 
In what follows, we will need the quantity $\Gamma^\omega(0,0,\theta,\theta')\equiv \Gamma^\omega(\theta-\theta')$.

%In the above expression, Eq.(\ref{eq:pi_interacting}), we actually neglected ($\omega,\,q$) dependence of the scalar vertex. 
The scalar vertex $\mathcal{T}^\omega$ 
%on the Fermi surface with small
takes on a simple form on the Fermi surface. For small external frequency and momentum values $\omega,\,v_Fq\ll\varepsilon_F$ the vertex can be decomposed as
\be\label{eq:T_expanded}
\mathcal{T}^\omega(\omega,\vec q, \theta)=\mathcal{T}_0+\mathcal{T}_1\omega+\mathcal{T}_2 q\cos\theta+...
\ee
where $\mathcal{T}_0=Z^{-1}$ by virtue of Ward's identity.\cite{lifshitz_pitaevskii} The linear terms 
%might have been important, if assessed 
are potentially relevant, if judged by power counting. However these contributions drop out, because for external frequency $\omega$ and momentum $\vec q$, the expressions in question contain both $\mathcal{T}^\omega(\omega,\vec q,\theta)$ and $\mathcal{T}^\omega(-\omega,-\vec q,\theta)$. This leads to a cancellation of the terms linear in  $\omega$ and $\vec q$.

%In fact, only first two terms of these series are 
Continuing with the analysis, we note that only the first two terms of the series in Eq.(\ref{eq:P0+P1+P2}) are relevant for long-wavelength plasmons with $v_Fq\ll\omega\ll \varepsilon_F$. Anticipating the square root dependence for plasmon frequency vs. wavenumber, we expand in $qv/\omega$ to obtain
\bea\nonumber
\Pi_1=&&\oint\frac{d\theta}{2\pi}\mathcal{T}^\omega (\omega,\vec q,\theta)\frac{\nu Z^2 vq\cos\theta}{\omega-vq\cos\theta}\mathcal{T}^\omega (-\omega,-\vec q,\theta)
\\
\label{eq:P1}
&&=\nu Z^2\mathcal{T}_0^2\oint\frac{d\theta}{2\pi}\frac{(vq\cos\theta)^2}{\omega^2}
=\frac{\nu}{2}\frac{v^2q^2}{\omega^2}
,
\\
\nonumber
\Pi_2=&&\oint\frac{d\theta d\theta'}{(2\pi)^2}\mathcal{T}^\omega (\omega,\vec q,\theta)\frac{\nu Z^2vq\cos\theta}{\omega}\Gamma^\omega(\omega,\vec q,\theta,\theta')
\\\nonumber
&& \times \frac{\nu Z^2 vq\cos\theta'}{\omega}\mathcal{T}^\omega (-\omega,-\vec q,\theta')
\\\nonumber
&&=\nu^2Z^4\mathcal{T}_0^2\frac{v^2q^2}{\omega^2}\oint\frac{d\theta d\theta'}{(2\pi)^2}\Gamma^\omega(\theta-\theta')\cos\theta\cos\theta'
\\
&&
=\frac{\nu}{2}\frac{v^2q^2}{\omega^2}\nu Z^2\oint\frac{d\theta}{2\pi}\Gamma^\omega(\theta)\cos\theta
.
\eea
The terms $\Pi_{n\ge 3}$, expanded in $qv/\omega$, yield contributions which are higher order in $q$ . The same is true for contributions arising from expanding $\mathcal{T}^\omega$, $\Gamma^\omega$ in powers of $\vec q$ and $\omega$ (with the exception for potentially relevant linear terms $\mathcal{T}_1$, $\mathcal{T}_2$ in Eq.(\ref{eq:T_expanded}) which merely cancel out). These terms are therefore not essential in the long-wavelength limit.  

Combining all the above results for $\Pi_0$ and $\Pi_{n\ge1}$, we find the long-wavelength asymptotic behavior for the net polarization function:
\bea
&&
\Pi(\vec q,\omega)=\frac{1}{2}\nu(1+F_1)\frac{v_F^2q^2}{\omega^2},
\quad 
\\\label{F_1}
&&
F_1=\nu Z^2\oint\frac{d\theta}{2\pi}\Gamma^\omega(\theta)\cos\theta
% \equiv \hat{F}[\cos\theta]
.
\eea
The quantity $F_1$ also 
%represents an eigenvalue 
gives the eigenvalues of the integral operator $\hat F$ corresponding to eigenfunctions $\cos\theta$ and $\sin\theta$. We can therefore write $F_1\cos\theta=\hat F[\cos\theta]$, which gives a Fourier harmonic of the operator kernel identical to Eq.(\ref{F_1}).

Plasmon dispersion can now be obtained from the relation $1-V(q)\Pi(q,\omega)=0$, giving Eq.(\ref{eq:plasmon_dispersion0}). The effects of interaction 
%between fermions are ``encapsulated'' in 
are encoded in the quantity ${Y}$ which equals Fermi velocity $v_0$ in the absence of interactions and is renormalized to a different value in an interacting system.

We note a difference between the quantities $F_m$ used in the FL literature\cite{lifshitz_pitaevskii} and 
%those used here.  The distinction
those used here, which is manifest in {\it their sign}. The difference
arises due to the long-range character of the $1/r$ interaction. In our case the density-density interaction $F(\theta-\theta')$ accounts for the effects due to exchange correlation but not for the Hartree effects. The Hartree contribution is expressed through the $1/r$ interaction taken at the plasmon momentum $\vec q$, corresponding to the Feynman graphs which can be disconnected by cutting a single interaction line. These contributions are incorporated in the dynamically screened interaction, Eq.(\ref{eq:vareps_zero}),  and hence not included in the definition of $\Gamma^\omega$ above. 
%This difference manifests itself in the sign of the interaction parameters. 
In contrast, for Fermi liquids with short-range interactions, the Landau interactions describing density-density response are dominated by the Hartree effects. As a result, they have positive sign for weak repulsive interactions. In contrast, our $F_m$ are negative, since they are dominated by exchange effects. In particular, we expect $F_1<0$. The negative sign, expected from this general reasoning, is also borne out by a microscopic analysis at weak coupling, see below.

We also note an interesting analogy between the approach developed in this section and the analysis of  superconducting Fermi liquids by Larkin and Migdal,\cite{larkin_migdal} and Leggett.\cite{leggett} Refs.\onlinecite{larkin_migdal,leggett} were concerned with Fermi-liquid renormalization of the quantities such as superfluid density in a metal with BCS pairing. Their analysis focused on the current correlation function which determines the response of current to vector potential, and followed similar steps as in the above discussion of $\Pi(\vec q,\omega)$. The renormalization effects were expressed through a combination of FL parameters, featuring a cancellation for a system with a parabolic band.

\section{Density dependence from one-loop RG}
\label{sec2}
In this section we derive plasmon dispersion for a simple model describing strongly interacting Dirac particles. This is done by employing  the renormalization group analysis developed in Refs.\cite{gonzalez1994,vafek2007,son2007}. 
%We treat the Coulomb interaction which mediates two-body scattering by accounting for dynamical screening in the random-phase approximation (RPA), giving 
We treat the two-body scattering vertex by accounting for dynamical screening of the Coulomb interaction in the random-phase approximation (RPA),
\be
%\label{eq:RPA}
U_{\vec q,\omega} = \frac{V(\vec q)}{\tilde{\kappa}(\vec q,\omega)}
%{1 - V_\vec{q}^0 \Pi (\omega, \vec q)}, \quad V_{\vec q}^0 = \frac{2\pi e^2}{ |\vec q|\kappa\varepsilon(\omega,\vec q}
,\quad
\tilde{\kappa}(\vec q,\omega)=1- V(\vec{q}) \Pi (\vec q, \omega)
.
\ee
Here 
%$V_{\vec q}^0 = 2\pi e^2/|\vec q|\kappa$ and 
the quantity $\tilde\kappa(\vec q,\omega)$ which describes dynamical screening is identical to that introduced in the above discussion of the dynamical density correlator, Eq.(\ref{eq:vareps_zero}).
Here $\Pi(\vec q, \omega)$ is the polarization function \cite{hwang1}
\be
\Pi (\vec q, \omega) =N\sum_{\vec k, s, s'} | F_{\vec k, \vec k + \vec q }|^2 \frac{ f(\epsilon_{\vec{k}, s}) - f(\epsilon_{{\vec{k}+\vec q}, s'})}{ i\omega + \epsilon_{\vec{k}, s}- \epsilon_{\vec{k} + \vec{q}, s'}  + i0}
,
\ee
with the band indices $\{s, s'\} = \pm $ and the coherence factors $| F_{\vec k, \vec k'}|^2=|\la \vec k',s'|\vec k,s\ra|^2$ describing overlaps of different pseudospin states. The polarization function is a sum of interband and intraband contributions, $\Pi=\Pi_1+\Pi_2$, described by $s'\ne s$ and $s'=s$, respectively. The factor $\epsilon(\vec q,\omega)$ describes the effect of intrinsic screening in graphene arising due to both the interband and intraband polarization. 
%Below we give expressions that will be used in our analysis (for a comprehensive treatment of the quantity $\Pi (\vec q, \omega)$ we refer to Ref.\onlinecite{hwang1}).

For undoped graphene, only interband transitions contribute, giving 
$\Pi_1  (\vec q, \omega)=-\frac{N}{16}\frac{ \vec q^2}{\sqrt{v^2\vec q^2+\omega^2}}$. This 
%dependence is key for 
expression is sufficient for our RG analysis [for a comprehensive treatment of the quantity $\Pi(\vec q,\omega)$ we refer to Ref.\onlinecite{hwang1}].\cite{gonzalez1994,vafek2007,son2007}

The full RG analysis of log-divergent corrections to Greens functions and vertices was performed in Refs.\onlinecite{gonzalez1994,vafek2007,son2007}. Below we use the results for one-loop RG calculation for large $N$. The 
%renormalization group flow for the velocity takes the form
RG flow for the quasiparticle velocity takes the form
\be
\frac{dv}{d\ell}=\beta v,\quad
\beta=\frac{8}{N\pi^2}
,
\ee
where $\ell=\ln(p_0/p)$ is the RG time parameter (here the  UV cutoff is set by interatomic spacing in graphene lattice, $p_0\sim a^{-1}$). This gives a power-law dependence 
\be\label{eq:v(p)}
v(p)=(p/p_0)^{-\beta} v_0
.
\ee
For $N=4$ we find  $\beta\approx 0.2$. This value is obtained from a one-loop RG which employs $1/N$ as a small parameter. The results for $N\sim 1$ are qualitatively similar, however the mathematical expressions are more cumbersome.  Acknowledging an approximate character of the scaling dimensions obtained from one-loop RG, we shall leave the exponent $\beta$ unspecified in the analytic expressions.

In the case of interest (doped graphene) the interband contribution $\Pi_1$ follows the above dependence for large momenta and frequencies, $|\vec q|\gg p_F$, $\omega\gg E_F$, 
%which provide dominant contribution to the RG flow of various quantities. 
which dominate the RG flow. The intraband contribution $\Pi_2$ is much smaller than $\Pi_1$ at 
%large energies and momenta, however these two contributions are 
such $\vec q$ and $\omega$, with the two contributions becoming comparable for $|\vec q|\sim p_F$, $\omega\sim E_F$. In the static limit, $\omega\ll E_F$, the polarization is dominated by the $\Pi_2$ contribution. In the range $q<2p_F$, which is where we will need it below, it is identical to that for two-dimensional systems with parabolic band,
% this contribution takes the form \addLL{[Check. Result for $q<2p_F$ identical to that for two-dimensional systems with parabolic band]}
\be
\label{eq:Pi2}
\Pi  (|\vec q|<2p_F)=-N\nu,
%\quad
%\Pi_2  (|\vec q|>2p_F)=-N\nu\frac{2p_F}{|\vec q|}
%-\frac{N}{16}\frac{ \vec q^2}{\sqrt{v^2\vec q^2+\omega^2}}
\ee
$\nu=p_F/2\pi v$ (we refer to Ref.\onlinecite{hwang1} for the analysis of other regimes).
%transitions contribute, giving $\Pi  (\vec q, \omega)=-\frac{N}{16}\frac{ \vec q^2}{\sqrt{v^2\vec q^2+\omega^2}}$. 
This gives a standard expression for the static RPA-screened interaction
\be\label{eq:U_q0}
U_{\vec q,0}=\frac{2\pi e^2}{\kappa|\vec q|+2\pi N\nu e^2}
\ee
We can obtain the two-particle scattering vertex $\Gamma^\omega$ by taking the interaction on the Fermi surface, $\vec q\sim p_F$, $\omega\ll E_F$. This gives
\be\label{eq:Gamma_RPA}
\Gamma^\omega(\theta,\theta')= - g_{\vec p,\vec p'}[{\cal T}(\Delta \vec p)]^2 U_{\Delta \vec p,0}
,\quad
\Delta p=2p_F\sin\frac{\Delta\theta}2
,
\ee
% where $\Delta p=2p_F\sin(\Delta\theta/2)$ and 
where $g_{\vec p,\vec p'}=|\la \vec p'\alpha'|\vec p\alpha \ra|^2=\cos^2(\Delta\theta/2)$ is the coherence factor describing the overlap of (pseudo)spinors describing quasiparticles at different points of the Fermi surface (here $\Delta\theta=\theta-\theta'$). The minus sign in Eq.(\ref{eq:Gamma_RPA}) arises because this expression represents a contribution from an exchange part of the two-particle vertex.\cite{lifshitz_pitaevskii}
%, see Ref.\onlinecite{lifshitz_pitaevskii}.

%Landau parameters can now be obtained from the relation between  the Landau functional and the vertex 
The FL interaction can now be obtained from its relation with the vertex
$\Gamma^\omega$, Eq.(\ref{eq:F=Gamma}). Combining  Eq.(\ref{eq:F=Gamma}) and  Eq.(\ref{eq:Gamma_RPA}), we find
%The form of $F(\theta-\theta')$ can be obtained from many-body calculation. For example, for weakly interacting Dirac particles we will obtain a relation \addLL{will or will not? sign?}
%
\be\label{eq:F_definition}
F(\theta-\theta')=-\nu  g_{\vec p,\vec p'} Z^2[{\cal T}(\Delta \vec p)]^2U_{\Delta \vec p,0}
\ee
In the large $N$ limit, the static RPA-screened interaction  can be approximated as 
$U_{\vec q,\omega}\approx -\frac1{\Pi(\vec q,\omega)}=\frac1{N\nu}$,
where we take into account that $|\Delta p|<2p_F$. 

Both $Z$ and ${\cal T}$ flow under RG, however their product remains equal to unity because of the Ward identity. As a result, 
%the Landau function does not 
FL interactions do not
undergo a power-law renormalization. 
Starting from $Z(p){\cal T}(p)=1$, where both $Z(p)$ and ${\cal T}(p)$ are given by power laws drawn from RG, we set $p=p_F$. This gives
%, and approximate ${\cal T}_{\Delta \vec p}\approx {\cal T}(p_F)$. This gives
% the Landau function
% can be obtained by taking this expression on the Fermi surface, $\vec q\sim p_F$, $\omega\ll E_F$, giving
\be
F(\theta-\theta')=
% \nu Z^2\frac{{\cal T}^2}{N\nu}\cos^2\lp\frac{\theta-\theta'}2\rp=
-\frac1{N}\lp\frac{{\cal T}(\Delta \vec p)}{{\cal T}(p_F)}\rp^2\cos^2\lp\frac{\theta-\theta'}2\rp
.
%\frac{4\cos^2(\theta/2)}{\pi N|\sin(\theta/2)|}
%\Gamma^\omega(\theta)
\ee
%
% where we ignored an additional angle dependence originating from the ratio ${\cal T}_{\Delta \vec p}/{\cal T}(p_F)$. 
We therefore conclude, that up to a remnant dependence on $p_F$ which may arise in the angle dependence due to the ratio ${\cal T}(\Delta \vec p)/{\cal T}(p_F)$, the function $F$ does not flow under RG. 

The function $F(\theta-\theta')$ is essentially independent of doping, whereas the velocity has a power-law dependence on doping, $v\propto p_F^{-\beta}$, given by Eq.(\ref{eq:v(p)}) for $p=p_F$. Combining these results, we find a power law dependence for plasmon dispersion,
\be\label{eq:w(k)_large_N}
\omega^2=A|\vec q|
,\quad 
A=\frac{Ne^2 p_Fv(p_F)}{2\kappa\hbar^2}(1+F_1)\propto p_F^{1-\beta}.
\ee
%where $v$ vs. $p$ dependence is given by Eq.(\ref{eq:v(p)}). 
This result is valid for plasmons with long wavelengths, $q\ll p_F$. The predicted power-law dependence 
%of plasmon on carrier concentration holds for both large and small densities,
holds in a wide range of carrier densities, both large and small, except very near the neutrality point where spatial inhomogeneity and thermal broadening play a role. 

To conclude, plasmon renormalization results from competition of two effects: plasmons tend to stiffen due to RG-enhancement of velocity, and to soften due to the negative sign of $F_1$. However, since $F_1$ does not flow under RG, whereas velocity $v$ does, the net effect of interactions 
%leads to stiffening of
is to stiffen plasmon dispersion. 
The predicted dependence 
$A\propto n^{(1-\beta)/2}$
can be used for extracting the exponent $\beta$ from measurement results.

\section{Magnetoplasmon in a Fermi liquid}

Below we analyze plasmon dispersion using FL transport equations.
%Below we will treat plasmons using Landau's transport equations, presenting the derivation in a way that will help to make connection to the microscopic derivation in Sec.\ref{sec1}. 
We will first deal with plasmons in the absence of magnetic field, then proceed to add a $B$ field. Some of the relevant quantites, such the Landau FL interaction $F(\theta-\theta')$ has already been introduced and analyzed, 
%here we re-introduce and discuss them again in the framework of Landau FL transport equation.
here we discuss them again to make connection to the microscopic derivation in Sec.\ref{sec1}.

In a semiclassical picture, the main
%In Landau's semiclassical framework, the fundamental
effect 
dominating the Fermi-liquid behavior is forward scattering wherein the whole system of interacting particles acts as a refractive medium 
%which makes each quasiparticle energy a 
in which a quasiparticle energy is a
function of occupancies of other particles. This is described by so-called Landau functional,\cite{lifshitz_pitaevskii} 
\be
\delta\epsilon(\vec p)=\int \frac{d^2p}{(2\pi)^2} f(\vec p,\vec p')\delta n(\vec p',\vec r),
\ee 
where $\delta n(\vec r,t)$ accounts for deviation of quasiparticle distribution from equilibrium. 

Since deviation from equilibrium occurs in a narrow band of states near Fermi surface, it is convenient to write the Landau functional by setting $|\vec p|=|\vec p'|=p_F$ and parameterizing the Fermi surface by a unit vector $\vec n=\hat{\vec p}$. Introducing the dimensionless Landau interaction $F(\vec p,\vec p')=\nu f(\vec p,\vec p')$, where $\nu=p_F /(2\pi \hbar^2 v)$ is the density of states per flavor, we write
\be\label{eq:epsilon(p)}
\epsilon(\vec p,\delta n)=\epsilon_0(\vec p)+\int \frac{d\theta'}{2\pi} F(\vec p,\vec p')\delta \tilde n(\vec p',\vec r)
%\int \frac{d^2p}{(2\pi)^2} f(\vec p,\vec p')\delta n(\vec p',\vec r)
.
\ee
Here $\epsilon_0(\vec p)=v(p-p_F)$ is linearized quasiparticle energy, the angle $\theta'$ describes orientation of $\vec p'$, and $\delta \tilde n(\vec p)$ is obtained by integrating $\delta n(\vec p)$ along the Fermi surface normal. The expression (\ref{eq:epsilon(p)}) can be treated as a Hamiltonian of one quasiparticle moving in a selfconsistent field of other quasiparticles. Equations of motion can then be obtained from Hamiltonian formalism via $\p_t n=\{H,n\}$. This gives
\be\label{eq:FL_dynamics}
(\p_t+\vec v\nabla) \delta n(\vec p,\vec r,t)=-\vec v\nabla\hat F\delta n(\vec p,\vec r,t),
\ee
where $\hat F$ is the integral operator defined in Eq.(\ref{eq:epsilon(p)}).

In a system with rotational symmetry, such as graphene and 2DEG's, the functional $F$ depends only on the angle between $\vec p$ and $\vec p'$:
\be
\hat F \delta n(\theta)\to\oint\frac{d\theta'}{2\pi}F(\theta-\theta') \delta n(\theta')
%,\quad F(\theta-\theta')=\nu_{0,*}V(\vec p-\vec p').
\ee
This expression defines a hermitian operator in the space of functions on the Fermi surface with the inner product
\be\label{eq:innerproduct}
\la f_1(\theta)|f_2(\theta)\ra=\oint\frac{d\theta}{2\pi} f^*_1(\theta)|f_2(\theta).
\ee
The eigenvalues of $\hat F$ are simply given by the Fourier coefficients
\be
F_m=\overline{F(\theta)e^{-im\theta}}=\oint \frac{d\theta}{2\pi}F(\theta)e^{-im\theta}
.
\ee
The quantities $F_m$ 
%represent the Landau interaction constants of a 2D FL system. 
parametrize FL interactions of a 2D system.
% \addLL{refer to prev sections}

To describe plasmons,
%and relate their dispersion with the quantities $F_m$ 
we add to Eq.(\ref{eq:FL_dynamics}) a long range electric field arising due to oscillating charge density,
\be
\lp \p_t+\vec v\nabla(\hat 1+\hat F)\rp \delta n(\vec p,\vec r,t)+e\vec E\nabla_{\vec p}n_0(\vec p)=0
%,\quad
%\vec v\nabla\hat F\delta n(\vec p,\vec r,t),
\ee
where  $n_0(\vec p)$ is the equilibrium Fermi distribution. Here $\vec E=-\nabla\Phi$, where $\Phi(\vec r)$ is the potential
\[
\Phi(\vec r)=\sum_i\int d^2r'\oint \frac{d\theta'}{2\pi} \frac{e}{\kappa |\vec r-\vec r'|}\delta n(\theta',\vec r',t).
\] 
Here  the sum is taken over $N$ spin/valley flavors, and the dielectric constant $\kappa$ accounts for screening by substrate.
% ($N=4$ in graphene, $N=2$ in GaAs). 
Performing Fourier transform, $\delta n(\theta,\vec r,t)=\int\int \frac{d\omega d^2k}{(2\pi)^3}\delta n_{\omega,\vec q}(\theta)e^{-i\omega t+i\vec q\vec r}$, we arrive at an eigenvalue equation of the form identical to that found in Sec.\ref{sec1} by analyzing poles of the dynamical screening function,
% $\epsilon(\vec q,\omega$
% \addLL{Sort out $k$ vs. $q$. Label duplicity.}
% identical to that familiar from the random-phase approximation
%
\be\label{eq:RPA}
1-V(\vec q)\Pi(\vec q,\omega)=0
\ee
The quantity $\Pi(\vec q,\omega)$ is identical to that found above by summation of FL-type ladder graphs,
%V(\vec k)=\frac{2\pi e^2}{|\vec k|\kappa}
\be
%\label{eq:pi_interacting}
\Pi(\vec q,\omega)=
%\oint \frac{d\theta}{2\pi}\oint \frac{d\theta'}{2\pi}
N\nu \,\tr_\theta\, \lp \frac{1}{\omega-\vec q\vec v(1+\hat F)}\vec q\vec v\rp
\ee
where trace is taken with respect to the inner product defined by Eq.(\ref{eq:innerproduct}). 
% The quantity $V(\vec q)=2\pi e^2/(|\vec q|\kappa)$ describes microscopic Coulomb interaction with a dielectric constant $\kappa$ accounting for screening by substrate.

Plasmon dispersion in the long wavelength limit can be found by expanding in the ratio $v|\vec q|/\omega$. We obtain
\be
\Pi (\vec q,\omega)\approx\frac{\nu}{\omega^2}\la \vec q\vec v|(1+\hat F)|\vec q\vec v\ra
=\frac{\nu q^2v^2}{\omega^2}\la \cos\theta|(1+\hat F)|\cos\theta\ra
% =\frac{Np_F k^2}{4\pi\omega^2}v(1+F_1)
\ee
Expressing the angle-averaged quantity through the Fourier coefficient $F_1$ and using the relation $\nu=p_F/(2\pi v)$ we rewrite this result as
\be
\Pi (\vec q,\omega)=\frac{Np_F \vec q^2}{4\pi\omega^2}v(1+F_1)
\ee
Plugging this into Eq.(\ref{eq:RPA}) and restoring Planck's constant, we obtain the same expression for plasmon dispersion as above, see Eq.(\ref{eq:plasmon_dispersion0}).

This analysis can be easily generalized to a system in the presence of an external magnetic field. This is done by accounting for the Lorentz force in the $\nabla_{\vec p}n$ term:
\be
\lp \p_t+\vec v\nabla(\hat 1+\hat F)\rp \delta n(\vec p,\vec r,t)+\lp e\vec E
% \nabla_{\vec p} n_0(\vec p)
+\frac{e}{c}\tilde{\vec v}\times \vec B\rp \cdot\nabla_{\vec p}n=0
\ee
where the velocity $\tilde{\vec v}=\nabla_{\vec p}\epsilon(\vec p,\delta n)$ includes 
the contributions accounting for the distribution function change
%contributions due to the distribution function 
$\delta n(\vec p)$.
This equation can be linearized as above, $n(\vec p,\vec r,t)=n_0(\vec p)+\delta n(\vec p,\vec r,t)$. In doing so, particular caution must be taken with the Lorentz force term since it is affected by 
% During the linearization of this equation we have to be careful:
the FL interactions. Accounting for the term in the velocity $\tilde{\vec v}$ that depends on $\delta n(\vec p)$, we write
\bea
& \tilde{\vec v}(n_0+\delta n)&=\nabla_{\vec p}\varepsilon(n_0+\delta n)
\\
\nonumber
&& 
=\nabla_{\vec p}\varepsilon+\nabla_{\vec p}\hat F\delta n =\vec v+\hat F \nabla_{\vec p'} \delta n,
\eea
Here we used Eq.(\ref{eq:epsilon(p)}) 
%and on the last step performed integration by parts. 
, performing integration by parts in the last term.

Terms linear in $\delta n$ can arise both from $\nabla_{\vec p}n_0$ and $\tilde{\vec v}$.  Taking a solution in a plane wave form $\delta n(\vec r,\vec p,t)\propto e^{-i\omega t+i\vec q\vec r}\nabla_\varepsilon n_0(\vec p) f(\theta)$
%we have
where $\theta$ is the angle between $\vec q$ and $\vec v$, we have
\bea
\lp i\omega-ivq \cos\theta(1+\hat F)\rp f(\theta)-\frac{e}{c}\lp\vec v\times \vec B\rp\cdot\nabla_{\vec p}f(\theta)
\\
\nonumber
-\lp\hat F \nabla_{\vec p}f(\theta) \times\vec B\rp \cdot\vec v
=i\nu V(q) qv\cos\theta \oint\frac{d\theta'}{2\pi}f(\theta').
\eea
%
%where $\theta$ is the angle between $\vec q$ and $\vec v$. 
This equation can be simplified as follows:
\bea
&&\lp i\omega-ivq \cos\theta(1+\hat F)-\omega_B(1+\hat F)\partial_\theta\rp f(\theta)
\\
\nonumber
&&=i\nu V(q) qv\cos\theta \oint\frac{d\theta'}{2\pi}f(\theta'),\quad \omega_B=\frac{evB}{p_F c}
.
\eea
%
%This is 
This gives
an eigenvalue problem with $\omega$ a  spectral parameter and $f(\theta)$ an eigenfunction. 
Inverting the operator on the left hand side gives a self-consistency equation 
\[
%\frac{qv}{\omega}
\oint\frac{d\theta}{2\pi}\frac{i\nu V(q)qv}{\omega+i\omega_B\left(1+\hat F\right)\partial_\theta
-qv\cos\theta(1+\hat F)}\cos\theta =1,
\]
% Eventually we arrive at self-consistency equation and use it to extract m
where $\frac1{\omega+...}$ is a shorthand for operator inverse. 
Magnetoplasmon dispersion can be obtained via perturbation theory in the parameter $qv/\omega\ll 1$,  giving
% Eq.(\ref{eq:magnetoplasmon_dispersion})
% recovering Pythagoras theorem
%
\be
\label{eq:magnetoplasmon_FL}
\omega^2(\vec q)=(1+F_1)^2\omega_B^2+(1+F_1)\frac{1}{2}V(\vec q) \nu \vec q^2v^2
\ee
This analysis ignores Bernstein modes which appear for $q r_c \sim 1$, 
%with $r_c$ the cyclotron radius.\cite{bernstein}
where $r_c$ is the cyclotron radius.\cite{bernstein} The validity of Eq.(\ref{eq:magnetoplasmon_FL}) is henceforth limited to long wavelengths, \mbox{$q r_c \ll 1$}.
Using the notation ${Y}=(1+F_1)v$ we arrive at Eq.(\ref{eq:magnetoplasmon_dispersion}). 
% \addLL{Since the value $F_1$ is positive and $v$ is enhanced by interactions (see above), magnetoplasmon dispersion is stiffenned by interactions.} 
Magnetoplasmon dependence on interactions is therefore described by the parameter ${Y}$ identical to that found for plasmons at $B=0$. 

As discussed above, the density dependence of the quantities $v$ and $1+F_1$ can be linked to their flow under RG. The power-law RG flow of velocity leads to stiffening of plasmon dispersion, which overwhelms the effect of softening due to the negative sign of $F_1$.

\section{Comparison to systems with parabolic band dispersion}
\label{sec:galilean}

To put the above results in perspective, 
%we recall that 
we recall some important aspects of
long-wavelength plasmons in 2D electron systems with a parabolic band.
%such as those in GaAs quantum well structures, afford a simple classical description. Plasmon dispersion in such systems can be derived from classical equations of motion making use of 
Such plasmons afford a simple description in terms of classical equations of motion for 
collective ``center-of-mass'' variables describing oscillating charge density.\cite{giuliani2005} The result is expressed in a general form through  unrenormalized band mass and electron interaction as
\be\label{eq:plasmon_unrenormalized}
m \omega^2 = n V(q) q^2 
\ee
where $n$ is carrier density and $V(q)=\frac{2\pi e^2}{|q|\kappa}$ for two-dimensional systems. An identical result is found for the quantum problem, since Heisenberg evolution generates classical equations of motion for the operators corresponding to the center-of-mass variables describing collective charge dynamics.

The absence of renormalization of plasmon dispersion, Eq.(\ref{eq:plasmon_unrenormalized}), can be linked to Galilean invariance. In quantum systems, Galilean invariance is a symmetry of the Hamiltonian generated by the transformation $x'=x+vt$, $t'=t$. This symmetry, which holds for any system with parabolic band dispersion and instantaneous interactions, 
%ensures that the effects of interaction disappear, 
ensure a complete cancellation of the effects of interaction,
rendering plasmon dispersion unrenormalized.
As discussed above, the cancellation of Fermi-liquid corrections follows from the Fermi-liquid identity\cite{lifshitz_pitaevskii} which relates renormalized velocity with the quantity $F_1$,
\be\label{eq:v/v*}
{Y}=(1+F_1)v=v_0
%\frac{v_0}{v}=1+\overline{F(\theta-\theta')\cos(\theta-\theta')}
,
\ee
where $v_0=p_F/m$ is Fermi velocity for noninteracting particles.
Crucially, 
%this identity holds only when  the overall band structure is parabolic, including the scales 
the validity of this identity depends on the band structure being parabolic on the scales
$\epsilon\gtrsim E_F$ and $\epsilon\sim E_F$ 
%on which electron-electron interactions give rise to Fermi-liquid constants. 
%on which the interaction contribution to the FL constants arise.
which determine the FL interactions.

The relation between unrenormalized plasmon dispersion and Galilean symmetry also holds in the presence of a magnetic field, wherein gapless plasmons turn into gapped magnetoplasmons. The magnetoplasmon dispersion is 
$\omega_B^2(q)=\omega_0^2(q)+\omega_c^2$, where $\omega_0(q)$ is given by Eq.(\ref{eq:plasmon_unrenormalized}) and $\omega_c=eB/mc$ is unrenormalized cyclotron frequency. 
In this case, the absence of renormalization is guaranteed by Kohn's theorem.\cite{kohn's theorem} The Kohn's theorem is established by treating collective charge dynamics in magnetic field using the center-of-mass variables in complete analogy with the derivation of Eq.(\ref{eq:plasmon_unrenormalized}). Because of the Galilean invariance, Heisenberg equations of motion for the center-of-mass variables obey classical  dynamics with unrenormalized cyclotron frequency. 

Unrenormalized plasmon dispersion also arises in other space dimensions, with  $V(q)=\frac{4\pi e^2}{q^2\kappa}$ for 3D systems and $V(q)=\frac{2e^2}{\kappa}\ln\frac1{|q|a}$ for 1D systems.  In the latter case, plasmon dispersion matches that of charge modes in one-dimensional Luttinger liquids. We stress that, in a general Luttinger liquid framework, the effective interaction for 1D plasmons 
%does not have to be equal to 
is distinct from
the bare interaction. 
%This equality arises in Luttinger theory as an additional ingredient due to Galilean invariance of particles with parabolic band dispersion. 
Nevertheless, due to Galilean invariance, plasmon dispersion in a 1D system with parabolic bands is expressed through unrenormalized bare interaction.
As noted above, what matters here 
%is not the linear dispersion in a system linearized near the Fermi level, but the character of the overall band structure. 
is the character of the overall band structure rather than the linear dispersion in a system linearized near the Fermi points.

In contrast to systems with parabolic dispersion, plasmons in graphene are sensitive to interactions. 
This is so because Galilean invariance is a non-symmetry for particles with linear dispersion, and hence the absence of renormalization is not guaranteed by any general principles.  As a result, plasmons in graphene feature a nontrivial dependence  on interactions. As we have seen above, plasmon dispersion is expressed through the Fermi velocity value $v$ which is renormalized by the interaction effects, and also through the FL interaction via a factor $1+F_1$. 
% These quantities, as we have seen above, are directly related to plasmon dispersion. 
We parenthetically note that electronic spectrum of graphene bilayer, while featuring parabolic band dispersion, does not obey Galilean invariance. We therefore expect plasmon dispersion in a  bilayer to exhibit a full-fledged Fermi-liquid renormalization, similar to graphene monolayer.

%Renormalization of plasmon dispersion results in a non-classical dependence of plasmon frequency on carrier density. An $n^{1/4}$ power law was predicted for weakly interacting electrons.\cite{wunsch06,hwang1,polini2009,dassarma09} This dependence, however, is not immune to interactions and therefore is not universal. As we have seen above, a significant deviation from the $1/4$ power law dependence is expected, which essentially maps out theoretically predicted RG flow of the quantity ${Y}=(1+F_1)v$. In that, the momentum dependence parametrized by Fermi momentum $p_F$ in the RG flow is manifested through the dependence on doping. This gives a unique opportunity to employ plasmon resonance as a probe of the RG theory of interaction effects in graphene.
To summarize, renormalization of electron properties due to interactions results in a nonclassical dependence of plasmon frequency on carrier density. Using a nonperturbative approach based on the FL theory, we show that plasmon dispersion can be expressed through Landau FL interactions. Measurements of plasmon resonance can therefore be used to extract the interaction parameters in a model-free way, which is particularly useful for studying strongly interacting systems such as graphene. Our results indicate a significant deviation from the $n^{1/4}$ power law dependence predicted for weakly interacting electrons in Refs.[\onlinecite{wunsch06,hwang1,polini2009,dassarma09}]. The density dependence predicted by our approach derives from the RG flow of the quantity $Y=(1+F_1)v$, where the RG ``time'' parameter value tracks the Fermi momentum. As an illustration, we consider RG for a large number of fermion flavors, which yields a power law of the form $n^{(1-\beta)/4}$, $\beta>0$. The density dependence of the plasmon resonance can therefore provide a direct, 
model-free probe of the RG theory of interaction effects in graphene.

We thank A. V. Chaplik, M. I. Dyakonov, F. H. L. Koppens, I. V. Kukushkin and M. Yu. Reizer for useful discussions. This work was supported in part under the MIT Skoltech Initiative, a collaboration between the Skolkovo Institute of Science and Technology (Skoltech), the Skolkovo foundation, and the Massachusetts Institute of Technology.

\end{document}